# Enhancing Capstone Program Workflow: A Case Study on a Platform for Managing Academic-Industry Projects

Luciano Pereira Soares[1], Rafael Corsi Ferrão[1]

[1] Insper, São Paulo, Brazil

Email: lpsoares@insper.edu.br, rafael.corsi@insper.edu.br

**Abstract**

Capstone projects are widely adopted by universities around the world as a culminating assessment in bachelor's degree programs. These projects typically involve student teams tackling complex, real-world problems proposed by external stakeholders, such as companies, NGOs, or research centers. Although they offer valuable hands-on experience, managing Capstone projects can be challenging due to their multiple stages and demands. The process typically begins by identifying students' interests, followed by sourcing and selecting potential projects from external organizations. After presenting these options to students, groups must be formed based on various criteria, including academic ranking, GPA, previous experience, and individual skill sets. In this paper, we detail a web-based tool designed to streamline the management of Capstone projects at Insper, with an emphasis on project sourcing and group formation. We also discuss the technological solutions, and the challenges encountered throughout development and deployment. Furthermore, we present usage data from recent years, offering insights that may prove valuable for institutions or teams developing similar tools in the future.

## 1  Introduction

A Capstone course typically requires students to develop a complex project in the final years of a bachelor's program, serving as a comprehensive academic and intellectual milestone that demonstrates their readiness for professional practice. They also often mark the first extensive endeavor that students embark on without the constraints of a predefined topic (Tenhunen *et al.*, 2023).

Due to the intensity and complexity of Capstone projects, students are often provided the opportunity to choose a project or topic that aligns with their interests, thereby boosting their intrinsic motivation. Additionally, certain group formation dynamics may be involved. However, this autonomy alone is insufficient to sustain engagement. Continuous assessments, whether individual or group-based, are essential throughout the project's development to maintain motivation and ensure consistent progress (Bessette, Okafor and Morkos, 2014).

Several institutions may also choose to undertake projects for external organizations to better prepare students for the job market (N and G, 2018). Securing high-quality projects that meet academic standards is a significant challenge, especially when projects are multidisciplinary and need to satisfy the requirements of multiple programs.

To ensure everything runs smoothly, groups typically have an advisor who orientates project dynamics, maintains a healthy relationship with the partner organization that proposed the project, addresses both technical and non-technical questions from students, reviews the project documentation, and frequently takes part in the assessment process (Paretti *et al.*, 2013).

Managing these aspects is highly complex, and the difficulty can quickly escalate depending on the number of students, projects, and the level of detail required in assessments. In this paper, we describe a pipeline and a platform developed to manage our specific style of Capstone projects.

This paper will focus on the stages associated with project sourcing, group formation, and project completion with the partnering organization, which correspond to capabilities 1, 2, 3, and 7 listed above.

## 1.1 Related Work

Due to the complexity of managing Capstone projects, both the academic community and educators have explored and developed various solutions to address specific challenges such as student planning, learning objective management, and individual evaluations. Many of these tools aim to enforce an agile methodology. For instance, the study by Umphress, Hendrix, and Cross (2002) analyzed 44 Capstone projects and concluded that "agile processes provide a bridge from ad hoc programming assignments to organized project work, and that process requires a suitable infrastructure of tools and process expertise." To facilitate the management of Capstone groups, several studies suggest using professional or custom software programs to assist in assessing and organizing teams (Olarte et al., 2014a; Fan, 2018).

Moreover, the application of tools in Capstone projects extends beyond group organization, as alternative methodologies have also been implemented and tested. Souza (2019) describes the development of a tool designed to gamify projects, while Neyem et al. (2025) propose using data extracted from Git repositories for individual evaluations and to monitor each group member's progress, providing evaluators with greater granularity. Additionally, given the challenges associated with forming groups in Capstone projects, specific tools have been developed to streamline group formation (Lan and Ginige, 2008).

Due to the inherent complexity, diverse methodologies, and unique characteristics of each Capstone project, institutions and educators have introduced integrated tools to manage the various demands more efficiently. Fan (2016) proposed a web-based tool that integrates project management and facilitates the submission of requirements by mentors, generating automated feedback for all parties involved. Similarly, Olarte et al. (2014b) introduced an integrated tool that simplifies monitoring the different phases of project development (planning, analysis, design, and production). In a comparable approach, Li et al. (2019) developed another tool that enhances feedback and phase management in IT Capstone projects. Lastly, Gan and Ouh (2020) detailed a web-based tool that supports comprehensive Capstone project management, including class and submission calendars, as well as the administration of project deliveries based on competencies and areas of interest.

## 2 Context

The open source tool (https://github.com/insper-education/pfe) was initially developed to manage Capstone projects at Insper - Brazil for computer, mechanical, and mechatronics engineering programs, and it has recently evolved to include students from the computer science program. At our institution, Capstone projects always address real-world challenges presented by external partners, such as companies, research centers, technology development organizations, or NGOs. Each project spans one semester and accommodates up to four students, who may come from different academic programs. Although, in theory, up to 200 students or more can be registered simultaneously, recent semesters have seen around 100 students working on 25 projects (Soares, 2024).

Capstone projects take place in the final years of a student's programs. In our case, for engineering students, this occurs in the second semester of the fourth year, while for computer science students, it happens in the second semester of the third year. By the time students begin their Capstone projects, they have completed all the required courses and are enrolled only in elective classes.

Students are expected to commit 24 hours per week to their Capstone projects, which typically last around four months. Each project is supervised by an advisor, a faculty member who dedicates 2 hours per week to assist the group, and is supported by a company representative with whom the students report weekly on their progress.

Evaluation methods are both individual and group-based. For individual assessments, students submit a biweekly self-report detailing their contributions and receive peer evaluations twice per project. These scores are used by the advisor to assess each student on four criteria: Technical Execution, Organization, Design, and Entrepreneurial Attitude. Group evaluations are conducted in three major phases: a preliminary report that outlines the project's objectives, followed by intermediate and final reports. The latter two are evaluated by a panel consisting of two external members (typically professors not directly involved in the project), ensuring that the group is assessed collectively using the same criteria as for all the projects

## 3  Process

The process begins with an initial meeting where students indicate their areas of interest, followed by planning, sourcing, and validation of project proposals from invited companies. Students then receive a list of approved projects and have the opportunity to ask questions before submitting their preferences. After groups are formed and advisors assigned, companies are notified, and official project work begins alongside academic activities. Midway and final reviews with the companies ensure progress, and the process concludes with an executive presentation and a request for publication authorization. The flow is detailed by day on Figure 1.

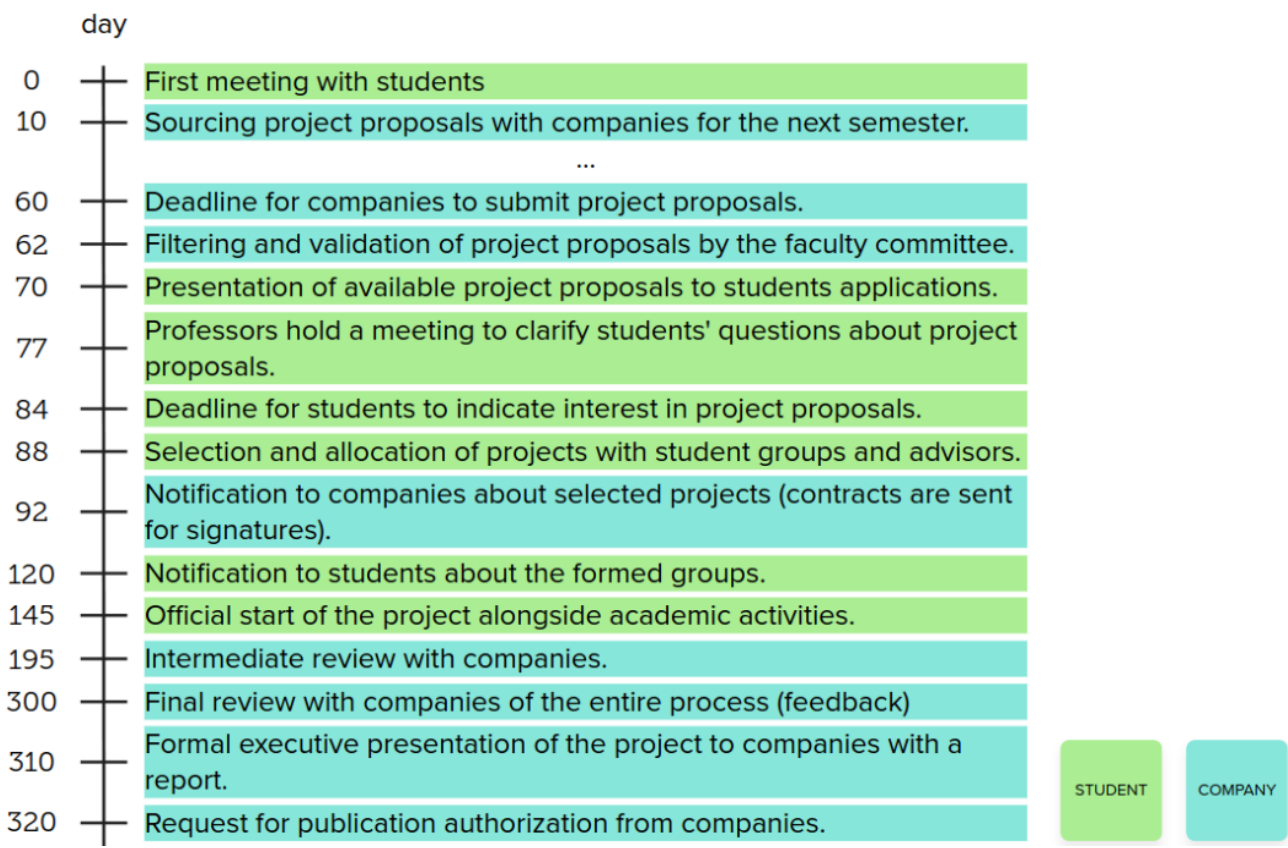

Figure 1. Capstone Project Timeline: This diagram outlines key milestones from the initial student meeting through proposal submission, project assignment, reviews, and concluding with the final executive presentation and publication authorization

## 3.1 Student Interests and Experiences

Once students are registered on the Capstone platform, they have access to a system that allows them to inform their areas of interest. This is one of the first actions they must take. These areas of interest will later be used to help the Capstone administration team to find projects in their preferred fields and to match the alignment between the projects chosen and their stated interests. A predefined table is provided to students, listing several areas: Computational Simulation; Industrial Automation; Embedded Systems; Data Science; Social Innovation; Interactive Systems; Information Systems; Bioengineering; Mobility Engineering; Dynamic Systems Control; Advanced Manufacturing; Administration, Economics, and Finance; Energy Efficiency; Robotics; Cloud Computing, that they have to tick. However, students are also encouraged to fill in an additional optional field if they have interests in other areas not included in the list.

In addition to selecting their areas of interest, students must answer whether they have worked, are currently working, have interned, or are currently interning at any company. This serves two purposes: first, to assess their professional experience, and second, to identify potential conflicts of interest if they are currently employed. Similarly, students are asked whether they have family members working at any company submitting a project proposal or at a direct competitor. This helps in identifying possible conflicts of interest.

Another experience-related question asks whether the student has participated in any student organizations, research programs, or other academic initiatives. This information can be used in the student selection and team formation process. Additionally, students are asked whether they have participated in social activities, such as volunteering or community work. This information is particularly relevant for projects with a social impact focus. A field for the student's LinkedIn profile is also available and can be used to review their professional background.

## 3.2 Organizations Outreach

Once the students' interests for the semester are known, contacts are made to partners to request project proposals. Organizations must fill out a form identifying themselves and describing their proposed project. The key fields in this form include:

- **Provisional Title of the Project**: A title that will be presented to students. This can be revised throughout the project.
- **Description of the Project**: A detailed proposal to be reviewed by professors and viewed by students.
- **Expected Results/Deliverables**: Examples include prototypes, models, tools, studies, validations, tests, or standards.
- **Resources Provided to the Student Team**: Organizations can indicate if they will provide any resources, such as development kits, raw materials, or access to cloud infrastructure (optional).
- **Other Relevant Observations**: Any additional information companies may wish to share with students (optional).

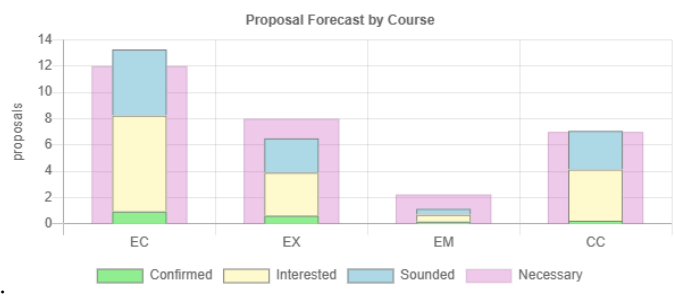

.

Figure 2. Graphical extract from the platform for the current semester, showing the number of students enrolled in Computer Engineering (EC), Computer Mechatronics (EX), Computer Mechanics (EM), and Computer Science. The "Necessary" box indicates the ideal number of projects required for each program.

The form also includes a section for companies to specify the areas involved in the project, similar to the student interest selection. As companies are invited to submit projects, an analysis is conducted to ensure a balanced number (neither too few nor too many). Institutions are categorized based on the types of projects they are expected to propose. Graphs, as shown in Figure 2, illustrate how the number of projects aligns with each program major.

All project proposals are stored and first evaluated by a Capstone committee composed of professors. This committee ensures that the proposed challenges meet all expected requirements, in a proposal conformity criteria:

- Does the project require knowledge and skills acquired in previous years of the undergraduate program?
- Does the project require students to apply a systematic technical design process for a system, component, or process to meet specific needs?
- Does the project represent a real-world or industry experience formally requested by an external organization?
- Does the project provide an opportunity to incorporate relevant technical standards that can be referenced in project reports?
- Does the project incorporate multiple realistic constraints, such as economic, environmental, social, political, ethical, health and safety, manufacturability, and/or sustainability considerations?
- Does the project require experimentation and hands-on skills?
- Does the project allow teamwork among students from one or more undergraduate programs?
- Is the project complexity appropriate for one academic semester (approximately 4.5 months)?
- Does the project have sufficient complexity to ensure that each team member contributes approximately 360 hours, including class time, lab work, meetings, and independent study?
- Does the project have concrete and measurable goals?

Professors also analyze each project and define the expected student profile for each major (Figure 3). For example, one project may be more suited for mechanical engineering students, while another may be better suited for computer science students. This presented information serves as a guideline, but students are free to apply to any project. However, this information is considered during the team formation phase.

**Expected Student Profile**
Each row represents a vacancy that should ideally be filled by a student from the marked programs

| Students | Computer Engineering | Mechatronic Engineering | Mechanical Engineering | Computer Science |
|---|---|---|---|---|
| #1 | ☐ | ☐ | ☑ | ☐ |
| #2 | ☐ | ☐ | ☑ | ☐ |
| #3 | ☐ | ☑ | ☑ | ☐ |
| #4 | ☑ | ☑ | ☑ | ☑ |

Figure 3: Professors selecting programs per student seat more suitable

If necessary, companies may be contacted to adjust their proposals, but this is often challenging when the application phase is already nearing completion. Projects that pass the conformity phase are then analyzed based on students' areas of interest. If there are project proposal gaps in certain fields, additional institutions may be contacted to submit proposals in those areas. Finally, an analysis is conducted to ensure that the projects adequately meet student demand. Figure 4 shows how the final result is convenient for the semester with the projects required for each program major.

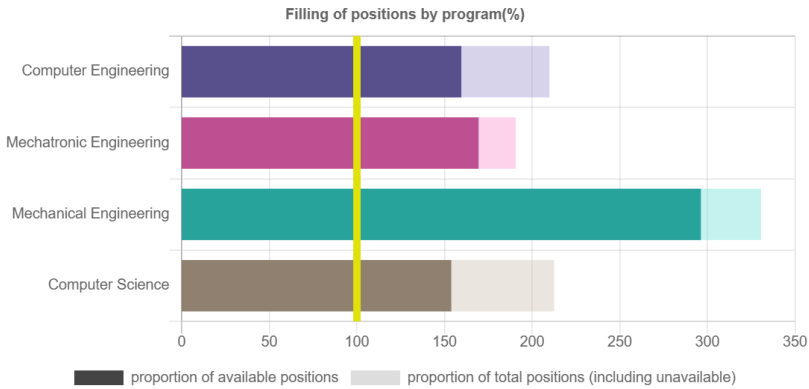

Figure 4: Each major with the approximate number of projects relative to demand. The yellow line represents 100% of the required student seats being filled. In this example, we can see that there are significantly more projects than needed to meet the required student seats.

## 3.3 Students Applying for Project Selection

Students must now prioritize the projects they are most interested in. In recent editions, they were required to select at least five preferred projects. They can review project details provided by companies, see which positions are recommended for their program, and check how many of their peers are interested in each project. Students have about two weeks to study the proposals and submit their applications. The image in Figure 5 shows that 7 students have already applied for this project proposal as their first option. When students see which projects are in high demand, interest tends to spread organically, simplifying the process of forming groups.

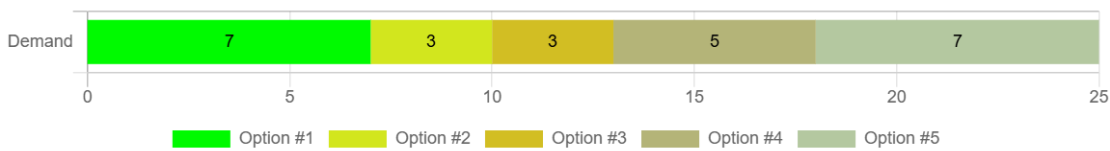

Figure 5: Example of information shows how other students are looking for a specific project

## 3.4 Group Formation

After students apply for projects, one of the most complex phases begins: forming the groups. Various experiments have been conducted to create the groups, but no objective function has been found that resolves all cases. Therefore, the adopted solution is to apply a heuristic that initially optimizes group formation based on the number of participants.

Groups with 2 or 3 students are prioritized since they have a higher probability of being completed with minor student adjustments. Students in groups with more than four members are analyzed to see if a redistribution can be made, moving them to groups with fewer than four members. This redistribution considers factors such as GPA and alignment of interest areas. However, the system is designed to avoid drastic changes to group compositions.

Once this phase is completed, a near-final solution is generated, typically very close to the ideal configuration for closing all groups. At this stage, students may be manually reassigned, considering more detailed information from their experience fields. Another manual step involves identifying potential conflicts of interest.

If a conflict is detected, a thorough analysis is conducted with all parties involved to determine whether the issue is significant or if it can be disregarded.

Once the groups are finalized, the process moves to the next phase, where advisors are assigned and each group must have a designated advisor. Finally, institutional partners are identified as the primary points of contact for mentorship activities and contractual matters.

## 3.5 Project execution

Project execution spans approximately four and a half months, guided by a faculty advisor who evaluates all interim deliverables and coordinates evaluation panels composed of two additional professors. Throughout the process, students must submit a biweekly individual report, as well as two activity reports (one individual and one group-based). For the final submission, the group—together with the advisor—can choose from three formats: a written report, a scientific paper, or a website.

Students are encouraged to meet weekly with a company representative to share progress and discuss results. However, the evaluation itself is entirely academic; the company does not take part in assigning grades. Each student receives one grade based on the group's deliverables and another reflecting their individual performance.

## 3.6 Project conclusion

The project concludes with the submission of a final report to the company and a formal presentation of the students' results, preferably at the company's site. Additionally, an in-person fair is held where students showcase their projects on posters; however, this event is not part of the evaluation process.

### 3.6.1 Overall Satisfaction

During the semester, information is also gathered from students and partners to help identify any issues with the projects. At the end of the project, main forms are distributed to both the partner organization and the students.

**Questions for the partners:**

Did the project progress as expected? If **YES**, do you know what the success factors were? If **NO**, can you identify the reasons? For example: delays in finalizing the scope, students not being very engaged, students having basic doubts about certain topics, or a team that wasn't fully committed.

One of the expectations for this activity was that students would gain a better understanding of the proposed challenge and discuss with you how to prototype and implement the agreed-upon solutions. Were students active in communicating, understanding your (and your end-users', if applicable) needs, and discussing possible solutions? Could you briefly comment on how these interactions went?

Was the student team allocated to your project properly organized? For example, were they punctual for meetings, did they behave appropriately, did they all dedicate themselves accordingly, and did they present an adequate proposal for conducting the project?

Do you have any other observations you would like to share with us? For example, regarding the interaction process with Insper, or something that Insper should have considered in this partnership proposal?

On a scale of 0 to 10, how much would you recommend the Capstone program to colleagues or partners from other companies?

**Questions for the students:**

1. How much would you recommend doing more projects in the upcoming semesters with the company where you worked on your project this semester?
2. Now that you know more about the company where you worked on your project this semester, would it be one of your top choices for an internship or permanent job?
3. Do you have any other comments you would like to share with us? This space is open for your observations, and we appreciate any detailed feedback you can provide.

## 4   Conclusion and Discussion

The Capstone program can be quite complex, involving many people from different areas. In our case, it currently includes around 20 advisor professors, 50 professors on examination boards, 100 students per semester, 100 company contacts per semester (both for project prospecting and active projects), two dedicated staff teams (one for prospecting and another for semester operations), and one main coordinator.

Over the past few years, our partner organizations have provided us with considerable feedback, consistently affirming that our projects not only meet but often exceed their expectations. With an average Net Promoter Score (NPS) of 9.45, it's clear that our partners are highly satisfied. Many are eager to submit new project proposals for the upcoming semester, whether by enhancing current initiatives or by introducing entirely innovative ideas. On the students' side, 77% strongly recommend the company, while 20% recommend it with reservations.

The tool not only provides control over all activities and ensures synchronization among all stakeholders, but it also enables the continuous improvement of the Capstone itself. Without it, managing these processes would demand significant time from everyone involved in the operation.

Additionally, this tool serves as an essential repository of academic information, helping faculty understand the impact of decisions made in a given semester. All reports are easily accessible, grades can be tracked, and various other convenient features are available. We extract data to use at ABET, ensuring that our processes meet accreditation standards while demonstrating continuous improvement in our curriculum. The system's robust analytical capabilities allow us to identify trends, assess learning outcomes, and adapt our instructional strategies accordingly. This not only supports our strategic planning and quality assurance efforts but also reinforces our commitment to academic excellence and the overall success of our students.